\documentstyle[12pt,aaspp4,psfig]{article}

\begin{document}

\def\b12{{B_{12}}}
\def\omegax{{\Omega_{\gamma}}}
\def\lgam{{L_{\gamma} }}
\def\fgam{{F_{\gamma} }}
\def\epsgam{{\epsilon_{\gamma}}}
\def\i45{{I_{45}}}

\def\be{\begin{eqnarray}}
\def\ee{\end{eqnarray}}

\title{The Gamma Ray Pulsar Population}

\author{M. A. McLaughlin\altaffilmark{1} \& J. M. Cordes\altaffilmark{1}}
\altaffiltext{1}{Astronomy Department, Cornell University, Ithaca, NY 14853}

\begin{abstract}
We apply a likelihood analysis to pulsar detections, pulsar upper limits, 
and diffuse background measurements from the OSSE and EGRET instruments
on the Compton Gamma Ray Observatory to
constrain the luminosity law for gamma-ray pulsars and some properties of the
gamma-ray pulsar population. We find that the dependence of luminosity
on spin period and dipole magnetic field is much steeper at OSSE
 than at EGRET energies (50$-$200 keV and $>$100 MeV, respectively),
suggesting that different emission
mechanisms are responsible for low- and high-energy gamma-ray emission. 
Incorporating a spin-down model and assuming a pulsar spatial distribution, we
estimate the fraction of the Galactic gamma-ray background due to 
unidentified pulsars
and find that pulsars may be an important component of the OSSE diffuse
flux, but are most likely not important at EGRET energies.
Using measurements of the diffuse background flux from these instruments, we are able
to place constraints on the braking index, initial spin period, and magnetic field of
the Galactic pulsar population, and are also able to constrain the pulsar birthrate to be
between 1/(25 yr) and 1/(500 yr). Our results are based on a large gamma-ray beam,
but they do not
scale in a simple way with beam size.
  We  
estimate that about 20 of the 169 unidentified EGRET sources are probably
gamma-ray pulsars. 
We use our model to predict the pulsar population that
will be seen by future gamma-ray instruments and estimate that GLAST will detect
roughly 750 gamma-ray pulsars as steady sources,
only 120 of which are currently known radio
pulsars.
\end{abstract}

\section{Introduction}

Since the launch of the Compton Gamma Ray Observatory (CGRO) in 1991,
 pulsed gamma-rays have been detected from
only 8 spin-driven pulsars. Because of the small number of detected pulsars, 
there remain many unanswered questions in pulsar gamma-ray astronomy.
The relationship between pulsar radio and gamma-ray emission is not clear. 
Aside from the Crab pulsar, the shapes of radio and gamma-ray profiles are quite
different, with the peaks of the radio and 
gamma-ray profiles falling at different pulse phases,
suggesting that low- and high-energy emissions
come from different regions of the magnetosphere. In fact,  Geminga,
an extremely
bright gamma-ray pulsar (GRP), is weak or quiet in the radio band. 
Furthermore, the gamma-ray profiles are broader than the radio profiles, suggesting that
the beaming solid angle for gamma-rays is larger than that for radio waves.
Models for the GRP emission mechanism, such as the polar cap and outer gap models,
have been invoked to 
explain the features of
GRPs, but no model has been successful
at explaining all of the observations.
 
Understanding GRPs may be the key to determining the
nature of the unidentified EGRET sources, perhaps the largest mystery remaining from
the CGRO mission.
Of the 271 sources listed in the Third EGRET Catalog (\cite{hartman99}),
169 are not identified as either pulsars or blazars, the
only known EGRET discrete sources.
While some of these sources are distributed isotropically and therefore
expected to be AGN, 73 of the 169 unidentified sources 
are located within $10^{\circ}$
of the Galactic plane. Since
pulsars are the only known Galactic EGRET sources, it
is likely that these unidentified sources are pulsars or an as-yet-unrecognized
source population. 
Modeling the GRP
population is important for estimating how many of the unidentified sources we 
expect to be pulsars, which of the unidentified sources are most likely to be pulsars,
and how we should go about detecting their pulsed emission.

Understanding GRPs may also be important for explaining the intensity, Galactic distribution, and
spectrum of the 
diffuse Galactic gamma-ray emission, the dominant feature of the gamma-ray sky.
 Accurate models for this diffuse emission are
essential for detecting
all but the very strongest gamma-ray sources. However,
it is not clear whether this emission is
due solely to cosmic ray interactions with interstellar material or partly
due to unidentified point sources. Modeling the GRP population allows us to
estimate the pulsar contribution to the diffuse emission. Furthermore, we can use
measurements
of the diffuse gamma-ray background to constrain important properties 
of GRPs.

As the CGRO mission is ending and future gamma-ray missions are being planned, it is
an appropriate time to assess the population characteristics of GRPs.
Suitable modeling allows us to constrain luminosity laws, luminosity evolution, and
spatial distributions
of GRPs and to
predict the pulsar population that future high-energy telescopes such as the Gamma Ray
Large Area Space Telescope (GLAST), with an estimated launch date of 2005,
will detect.

In order to address the above issues within the confines of the small number of detected
GRPs, we
have developed a likelihood analysis which employs all
available information about GRPs, including pulsar detections, pulsar upper limits, and
diffuse background measurements, to constrain their luminosity law and 
population parameters.
In this paper, we describe the included data from the CGRO instruments
OSSE (Oriented
Scintillation Spectrometer Experiment) and EGRET
(Energetic Gamma Ray Experiment Telescope).
We present our model for pulsar gamma-ray luminosity and luminosity
evolution, explain our likelihood analysis method, and discuss our results and
their implications.

\section{The Data}
\subsection{OSSE}
Three pulsars (the Crab and Vela pulsars, and PSR B1509$-$58)
have been detected by OSSE
(\cite{ulmer94}; \cite{matz94}; \cite{strickman96}).
Table 1 lists their 50$-$200 keV fluxes,
along with upper and lower distance bounds,
spin periods, period derivatives,
surface dipole magnetic fields (i.e. $B_{12} \equiv \sqrt{10^{15}P\dot{P}}$, for $P$ in seconds and $\b12$ in units of $10^{12}$ 
Gauss),
and total spin-down energy loss rate (i.e. $\dot{E} = I\Omega\dot{\Omega}$, where
$I$ is the moment of inertia\footnote{We assume $I = 10^{45}$ g cm$^{2}$ throughout.}
and $\Omega$ is the spin frequency). We calculate
the distance $D$ to each pulsar using the Taylor and Cordes (1993) model for
Galactic electron density, except for pulsars with more
accurate distance measurements from parallax or a supernova remnant association.
Following Taylor and Cordes, we
include a 25\% error in our distance estimation.
We also include
upper limits to the pulsed OSSE flux for 27 known pulsars
(\cite{sch95}; \cite{ulmer99}). We assume
a beaming solid angle $\Omega_{\gamma}$ of $2\pi$, consistent with
the uniformly wide pulse profiles observed for the detected pulsars.
A photon
spectrum with a power law index
of $-2$ (i.e. $F \propto \nu^{-2}$) has been assumed to convert all upper limits
to the 50$-$200 keV energy range. While measured spectral indices for B1509-58
(\cite{wilson93})
and Vela (\cite{strickman96}) are greater than $-2$, adopting a harder spectrum
for the conversion of upper limits to luminosity changes the derived upper limits by only a
few percent
and does not affect the results of this analysis.
We also make use of measurements of the OSSE diffuse
gamma-ray background at 3 locations in the Galactic plane.
Table 2 lists the central longitude $l$
of each of these pointings, the viewing periods used, and
the measured diffuse flux (\cite{skibo97}; \cite{kinzer99}). Viewing periods, or time intervals
over which the spacecraft pointing was fixed, are defined in the Third EGRET Catalog
(\cite{hartman99}). 
A power law spectrum with index $-2$ has been assumed to calculate the integrated 50$-$200
keV diffuse flux. Choosing a different spectral index does not significantly change 
results\footnote{For a spectral index of $-$1.5, total diffuse flux over the 50$-$ 200 keV band
is 1.1 times that in
Table 2. For a spectral index of $-$2.5, total diffuse flux is 0.9 times that given in
Table 2.}.

\begin{deluxetable}{lccccccc}
\tablewidth{7.0in}
\tablenum{1}
\tablecaption{OSSE Pulsar Detections}
\tablehead{
\colhead{Name} & \colhead{$F$} & \colhead{$D_{l}$} & \colhead{$D_{u}$} & \colhead{$P$}
 & \colhead{$\dot{P}$} & \colhead{$\b12$} & \colhead{$\dot{E}$} \\
 & \colhead{(ergs kpc$^{-2}$ s$^{-1}$)} &  \colhead{(kpc)} & \colhead{(kpc)} & \colhead{(ms)}
 & \colhead{(s/s)} & \colhead{($10^{12}$ G)} & \colhead{(ergs/s)}}
\startdata
B0531$+$21& $3.33 \pm 0.4 \times 10^{34}$ & 1.50 & 2.50 & 33.4& $ 4.21 \times 10^{-13}$& 3.8 & $4.47 \times 10^{38}$ \nl
B0833$-$45& $8.34 \pm 5.6 \times 10^{31}$ & 0.40 & 0.60 & 89.3 & $1.25 \times 10^{-13}$& 3.4& $ 6.92 \times 10^{36}$   \nl
B1509$-$58& $1.24 \pm  0.2 \times 10^{33}$ & 3.50 & 5.30 & 150.6  & $1.54 \times 10^{-12}$& 15.5 & $1.78 \times 10^{37}$ \nl
\enddata
\end{deluxetable}{}
\begin{deluxetable}{cccc}
\tablewidth{7.0in}
\tablenum{2}
\tablecaption{OSSE Diffuse Measurements at b = 0$^{\circ}$}
\tablehead{
\colhead{l} &  \colhead{VPs} & \colhead{$F$} \\
\colhead{(degrees)} &  &\colhead{(ergs kpc$^{-2}$ s$^{-1}$ sr$^{-1}$)}
}
\startdata
0 &  $5 + 16$ & $1.41 \pm 0.2 \times 10^{37}$ \nl
25 &  $7 + 13$ & $4.67 \pm 1.3 \times 10^{36}$ \nl
95 &  $515 + 519$ & $3.53 \pm 1.2 \times 10^{36}$ \nl
\enddata
\end{deluxetable}{}

\subsection{EGRET}

We include 7 EGRET pulsar detections in our analysis (\cite{nolan96}; \cite{kaspi99}).
Table 3 lists the E $>$ 100
MeV fluxes for these pulsars, along with distance bounds,
periods, period derivatives, 
magnetic fields, and spin-down energies.
We also include 353 upper limits to the pulsed flux for EGRET
pulsars (\cite{nel96}; \cite{fierro95}; \cite{arz97}).
Again, we assume
a beaming solid angle of 2$\pi$ and 
a photon power law index of $-2$.
Table 4 lists the included measurements of the EGRET diffuse gamma-ray
background (\cite{hunter97}) for E $>$ 100 MeV, calculated by assuming a spectrum
with a power law index of $-2$.
Choosing a different spectral index will change the derived diffuse fluxes\footnote{For
a spectral index of $-$1.5, diffuse fluxes are 3 times that listed
in Table 4. For a spectral index of $-$2.5, total diffuse fluxes are 0.5 times that listed in
the table.}; the impact of this on the analysis is discussed further in Section 5.2.
Individual viewing periods are not listed, as fluxes were calculated from an
EGRET composite map.

\begin{deluxetable}{lccccccc}
\tablewidth{7.0in}
\tablenum{3}
\tablecaption{EGRET Pulsar Detections}
\tablehead{
\colhead{Name} & \colhead{$F$} & \colhead{$D_{l}$} & \colhead{$D_{u}$} & \colhead{$P$} & \colhead
{$\dot{P}$} & \colhead{$\b12$} & \colhead{$\dot{E}$} \\
& \colhead{(ergs kpc$^{-2}$ s$^{-1}$)} & \colhead{(kpc)} & \colhead{(kpc)} & \colhead{(ms)} &
 \colhead{(s/s)} & \colhead{($10^{12}$ G)} & \colhead{(ergs/s)}}
\startdata
B0531$+$21& $9.55 \pm 0.96 \times 10^{33}$ & 1.50 & 2.50 & 33.4& $4.21 \times 10^{-13}$& 3.80 & $4.47 \times 10^{38}$  \nl
J0633$+$1746& $3.53 \pm 0.35 \times 10^{34}$ & 0.12 & 0.22 & 237.1& $1.10 \times 10^{-14}$& 1.62 & $3.24 \times 10^{34}$ \nl
B0833$-$45& $6.78 \pm 0.68 \times 10^{34}$ & 0.40 & 0.60 & 89.3 & $1.25 \times 10^{-13}$& 3.39   & $6.92 \times 10^{36}$ \nl
B1046$-$58&$2.39 \pm 0.57 \times 10^{33}$ & 2.24 & 3.73 & 123.7 & $9.59 \times 10^{-14}$& 3.47 & $ 2.00 \times 10^{36}$ \nl
B1055$-$52& $4.01 \pm 0.80 \times 10^{33}$ & 1.15 & 1.91 & 197.1 & $5.83 \times 10^{-15}$& 1.10 & $3.02 \times 10^{34}$ \nl
B1706$-$44& $7.92 \pm 1.59 \times 10^{33}$ & 1.37 & 2.28 & 102.4 & $9.30 \times 10^{-14}$& 3.09 & $3.39 \times 10^{36}$ \nl
B1951$+$32& $2.29 \pm 0.46 \times 10^{33}$ & 1.78 & 2.96 & 39.5 & $5.84 \times 10^{-15}$& 0.49 & $3.72 \times 10^{36}$ \nl
\enddata
\end{deluxetable}{}
\begin{deluxetable}{ccc}
\tablewidth{7.0in}
\tablenum{4}
\tablecaption{EGRET Diffuse Measurements at b = 0$^{\circ}$}
\tablehead{
\colhead{l} & \colhead{$F$} \\
\colhead{(degrees)} & \colhead{(ergs kpc$^{-2}$ s$^{-1}$ sr$^{-1}$)} }
\startdata
20 &  $1.70 \pm 0.30 \times 10^{41}$ \nl
40 &  $1.44 \pm 0.25 \times 10^{41}$ \nl
60 &  $7.85 \pm 0.40 \times 10^{40}$ \nl
\enddata
\end{deluxetable}{}

\section{Luminosity and Population Model}
\subsection{Luminosity and Period Model}
We model a pulsar's gamma-ray luminosity $L$ as a power law in period and magnetic field,
\be
L = \gamma P^{-\alpha}\b12^{\beta},
\label{eq:lummodel}
\ee
where $P$ is the period in seconds and $\b12$ is the surface dipole magnetic field
in
units of $10^{12}$ Gauss. If the luminosity predicted by Eq.~\ref{eq:lummodel} is greater than some fraction $\epsgam$
of the pulsar's total spin-down energy $\dot{E}$ (i.e.
$\dot{E} = I\Omega\dot{\Omega}
\propto \dot{P}P^{-3} \propto P^{-4}B^{2}$), the model luminosity
 is set equal to $\epsgam\dot{E}$ (see Section 5). 

Given a luminosity law and assuming a spindown law 
$\dot{\Omega} \propto \Omega^n$, where $\Omega$ is the spin frequency and $n$ is
the braking index,
we may calculate a 
population-averaged gamma-ray luminosity, from which the total diffuse
flux from GRPs may be derived.
 
The spin period at age $t$ of a pulsar born with initial 
spin period $P_0$ is
\be
P(t) = P_0 \left ( 1 + t/\tau_0 \right)^{1/(n-1)},
\ee
where $\tau_0 \equiv P_0/(n-1)\dot{P_0}$ is the initial spindown 
time.
We assume the magnetic field remains constant
throughout the pulsar's lifetime. Therefore, as $B_{12} \propto \sqrt{P\dot{P}}$ for $n = 3$,
$P_0\dot{P_0} = P\dot{P}$. However, while we expect $n=3$
 for magnetic dipole radiation, all measured
pulsar braking indices are less than 3 (\cite{lyne93}; \cite{kaspi94}; \cite{gouiffes92}).
 If $n$ does not equal 3, then the magnetic field estimate $\b12$
must be viewed as only an effective field strength that may differ substantially from
the actual dipole field strength.

For a given magnetic field strength $\b12$, the probability density function
(PDF) for period $P$ over a population of objects born at constant birth rate is
\be
f_P(P\vert \b12) = \left ( T_g \dot{P} \right)^{-1}, \quad P_0 \le P \le P_g,
\label{eq:perpdf}
\ee
where $T_g$ is the age of the Galaxy and $P_g \equiv P_0 (1 + T_g/\tau_0)^{1/(n-1)}$
 is the period of the oldest pulsar. We note that for fixed $\b12$, $\dot{P}$ is
uniquely determined by $P$. Throughout the analysis, we make the simplistic assumption
that the PDF for $\b12$ is a 
delta function to keep the number of fitting parameters small.

The total number of pulsars in the Galaxy is
\be
N_{psr} = \dot N_{psr} T_g = 10^8 \dot N_{0.01} T_{g,10},
\ee
for a birth rate $\dot N_{psr} = 0.01$ yr$^{-1} \dot N_{0.01}$
and Galactic age $T_g \equiv 10^{10} T_{g,10}$ yr.
The differential distribution of their spin periods is
\be
\frac{dN_{psr}}{dP} = N_{psr} f_P(P\vert \b12)
              = 2 \dot N_{psr} T_g P P_g^{-2}
              = 10^{8.3} {\rm \,\, s^{-1}} \dot N_{0.01} T_{g,10} P  P_g^{-2}.
\ee
Note that, because $P_g$ depends on $n$, this expression is dependent upon $n$.

The luminosity PDF for a given magnetic field strength $\b12$, calculated
from Eqs.~\ref{eq:lummodel} \&~\ref{eq:perpdf} is
\be
f_{L}(L \vert \b12)
    = \frac{f_P(P \vert \b12)} {\vert dL/dP \vert}
    = \frac{2\gamma^{2/\alpha}}{\alpha P_g^{2}} \b12^{2\beta/\alpha}
              L^{-(1+2/\alpha)},
\label{eq:lumpdf}
\ee
and the population-averaged GRP luminosity is therefore
\be
\langle L \vert \b12 \rangle =
     \frac{2 \gamma \b12^{\beta}
     P_0^{2 - \alpha}}{P_{g}^{2}(\alpha-2)}
     \left [1- \left (1 + \frac{T_{g}}{\tau_{0}} \right) ^{
     \left (\frac{2-\alpha}{n-1}\right)} \right ] \hspace {0.15in} \rm for \hspace{0.1in}  \alpha \ne  2 \hspace{0.1in} and \hspace{0.1in} n \ne 1
\label{eq:avglum}
\ee
\be
        \approx \frac{10^{15} \gamma \b12^{\beta - 2}
     P_0^{2 - \alpha}}{T_{g}(\alpha-2)}
        \hspace {0.15in} \rm for \hspace{0.1in}  \alpha > 2 \hspace{0.1in} and \hspace{0.1in} n = 3.
\ee

\subsection{Spatial Distribution}

We adopt a spatial distribution in order to
calculate the total gamma-ray flux from pulsars in some direction through the
Galaxy.
We assume that pulsars are distributed in a Gaussian disk of radius $r_g$ with
exponential scale height $h$. We also include a molecular ring of width $w_r$ at
Galactic radius $r_r$.
We can express the
pulsar number density at position ${\bf x}$ as a function of radius from the
Galactic center $r$ and
height
above the Galactic plane $z$ as
\be
\rho({\bf x}) = \rho(r,z) = \frac{1}{4\pi h\sqrt{\pi}(r_g+\eta w_r)} e^{-z/h}(e^{-r^2/r_g^2} + \eta e^{-(r-r_r)^{2}/w_r^2})\rho_0
\label{eq:spatial}
\ee
where $\eta$ is a dimensionless 
ring strength parameter and $\rho_0 \equiv N_{psr}/V_g$ is the
average Galactic pulsar number 
density, with $N_{psr}$ the number of pulsars in the Galaxy  and 
$V_g$ the total Galactic volume. This expression
is normalized according to $\int d^{3}{\bf x} \rho({\bf x}) = N_{psr}$. We also define
a dimensionless spatial distribution $G({\bf x})
\equiv \rho({\bf x}) V_g/N_{psr}$ such
that $\int d^{3}{\bf x} G({\bf x}) = V_g$. To convert the PDF in
$r$ and $z$ (Eq.~\ref{eq:spatial}) to a PDF $f_D(D)$ in distance from the Sun $D$,
we 
use the relation
\be
D = [r^2 + z^2 - 2R_{\odot}r\sin{\phi} + R_{\odot}^2]^{1/2}
\ee
where $r^2 = x^2 + y^2$ for $x = r\cos{\phi}$ and $y = r\sin{\phi}$, and
$R_{\odot}$ is the distance of the Sun from the Galactic center. Assuming cylindrical 
symmetry, 
we integrate over $\phi$ to find 
\be
f_D(D) = \frac{D}{2 \pi R_{\odot}}~\int\int dr~dz~\rho(r,z)~\frac{1}{r}~\left[1-\left(\frac{r^{2}+z^{2}+R_{\odot}^{2}-D^{2}}{2R_{\odot}r}\right)^{2}\right]^{-1/2}
\label{eq:distancepdf}
\ee
where $|r^{2}+z^{2}+R_{\odot}^{2}-D^{2}| \le 2R_{\odot}r$ for all $r$ and $z$. The integrals
over $r$ and $z$ are done numerically.

\subsection{Flux Distribution}

The PDF for the flux F for a given gamma-ray luminosity $L$ 
in terms of the distance PDF $f_D(D)$ is
\be
f_F(F\vert L) = \frac{f_D(D)} {\vert dF/dD \vert}
                = \frac{1}{2} \left ( \frac{L}{\omegax} \right )^{1/2} f_D(D) F^{-3/2}
\ee
for beaming solid angle $\omegax$.
The flux PDF after integrating over a luminosity PDF, for fixed $\b12$, is
\be
f_F(F\vert \b12) = \int dL~f_{L}(L\vert \b12) f_F(F\vert L).
\ee
Substituting Eq.~\ref{eq:lumpdf}, we do this integral numerically, with integration limits determined by
the luminosities corresponding to periods $P_0$ and $P_g$. If the luminosity predicted for
period $P_0$ is greater than $\epsgam\dot{E}$, the maximum luminosity is
set equal to $\epsgam\dot{E}$. In this regime,
$f_{L}(L \vert \b12)$ scales as $L^{-3/2}$ instead of the $L^{-(1+2/\alpha)}$ of Eq.~\ref{eq:lumpdf}. Including a
maximum allowed gamma-ray efficiency is therefore important for all models with $\alpha < 4$.

The differential number of pulsars vs. flux, with beaming solid angle $\Omega_{\gamma}$,
 is
\be
\frac{dN_{psr}}{dF} = \frac{\Omega_{\gamma}}{4\pi} N_{psr} f_F(F \vert \b12),
\ee
which can be expressed as a
histogram of $F$ with logarithmic bins of width $\Delta \log F$ 
\be
\Delta N_{psr}(F) = 2F \sinh(1.15 \Delta \log F ) \frac{dN_{psr}}{dF}.
\label{eq:hist}
\ee

Eq.~\ref{eq:hist} gives the expected flux distribution for all Galactic pulsars. To 
calculate the expected flux from all pulsars in one beam (i.e. a cone of directions encompassing
a small solid angle) centered on a particular 
direction, we take the average
luminosity from Eq.~\ref{eq:avglum} and integrate across the solid angle of the beam
and our dimensionless spatial distribution. The total flux due to pulsars
for a beam centered on Galactic longitude $l$ and latitude $b$ and of solid angle
$\Omega_{b}$ is
\be
F(l,b,\Omega_{b}) =  \frac{N_{psr}}{12\pi V_g} \langle L \vert \b12 \rangle \int d\Omega_{b}(l,b)\int dD~G({\bf x}),
\label{eq:beamflux}
\ee
where distance $D$ from the Sun is related to $r$ and $z$ of
 Eq.~\ref{eq:spatial} through
\be
x = D \cos{b} \sin{l} \\
y = R_{\odot} - D \cos{b}\cos{l} \\
z = D \sin{b}.
\ee

\section{Likelihood Analysis}

We use a likelihood analysis which exploits the detected GRPs,
GRP upper limits, and GRP diffuse background measurements to find the best model
to describe the GRP population. The total likelihood for model ${\bf \Theta}$
(i.e. one combination of parameters $\alpha$,
$\beta$, $\gamma$, $n$, $P_0$, and $\b12$) is factorable into the likelihoods for detections, upper limits,
and diffuse background measurements and can be expressed as 
\be
 {\cal L}_{\rm tot}({\bf \Theta}) = {\cal L}_{\rm det}({\bf \Theta})
{\cal L}_{\rm up}({\bf \Theta}) {\cal L}_{\rm dif}({\bf \Theta}).
\label{eq:totlike}
\ee
The total likelihoods for the detections, upper limits,
and diffuse measurements are the products over the individual measurements such that
\be
{\cal L}_{\rm det}({\bf \Theta}) = \prod_{i=1}^{N_{det}} {\cal L}_{det,i}({\bf \Theta}), \\
{\cal L}_{\rm up}({\bf \Theta}) = \prod_{i=1}^{N_{up}} {\cal L}_{up,i}({\bf \Theta}),~{\rm and} \\
{\cal L}_{\rm dif}({\bf \Theta}) = \prod_{i=1}^{N_{dif}} {\cal L}_{dif,i}({\bf \Theta}),
\ee
where $N_{det}$, $N_{up}$, and $N_{dif}$ are the number of included detections,
upper limits, and
diffuse pointings, respectively. 

\subsection{Detected Pulsars}

To calculate ${\cal L}_{det,i}({\bf \Theta})$, the likelihood of 
model ${\bf \Theta}$ given 
the $i$th pulsar detection, we use Eq.~\ref{eq:lummodel}
to calculate the predicted model
luminosity $L_i({\bf \Theta})$
for the pulsar's measured $P$ and $\dot{P}$ and for model parameters
$\alpha$, $\beta$, and $\gamma$. For an assumed pulsar distance $D$, we 
calculate a predicted flux $\hat{F_i}({\bf \Theta},D) =
L_i({\bf \Theta})/\Omega_{\gamma}D^{2}$ and compare this to the
detected flux $F_i$.
The form of the
likelihood function is chosen to be a Gaussian 
with width determined by the flux measurement error ${\sigma_{F}}_{i}$.
Integrating
over a distance PDF $f(D)$ (n.b. different from the distance PDF of Eq.~\ref{eq:distancepdf})
to account for uncertainties in distance estimation,
we may express
the PDF for model ${\bf \Theta}$ given detection $i$ as
\be
{\cal L}_{det,i}({\bf \Theta}) = \int_{D_l}^{D_u} dD~f(D)~g((\hat{F}_i({\bf \Theta},D) - F_i);{\sigma_{F}}_{i}),
\label{eq:detlike}
\ee
where $g(x,\sigma) = (2\pi\sigma^2)^{-1/2} {\rm exp}(-x^2/2\sigma^2)$.
The likelihood for a particular model 
maximizes when the pulsar's predicted flux $\hat{F_i}({\bf \Theta},D)$ equals the
detected
flux $F_i$.
We assume a uniform distance
distribution (i.e. $f(D) =
1/(D_u-D_l)$ for $D_l \le D \le D_u$ and
0 otherwise).

\subsection{Upper Limits}

For each pulsar with a measured upper limit, we calculate the predicted luminosity
given its measured $P$ and $\dot{P}$ and for
model parameters $\alpha$, $\beta$, and $\gamma$. For an assumed pulsar distance $D$, we 
calculate a model flux $\hat{F_i}({\bf \Theta},D)$ and compare this to the measured upper
limit $F_i$. 
The form of the upper limit likelihood
function is chosen to be flat with a
one-sided Gaussian roll-off with width determined by the upper limit 
measurement error ${\sigma_{F}}_{i}$.
Again integrating over a distance PDF $f(D)$, we express the PDF for model
${\bf \Theta}$ given upper limit measurement $i$ as
\be
{\cal L}_{up,i}({\bf \Theta}) = \int_{D_l}^{D_u} f(D)~dD~\frac{\int_{\hat{F}_i({\bf \Theta},D)}^{\infty} dF~g(F - F_i);{\sigma_{F}}_{i})}{\int_{F_i}^{\infty} dF~g(F - F_i);{\sigma_{F}}_{i})} 
\label{eq:uplike}
\ee
for $\hat{F_i}({\bf \Theta},D) \ge F_{i}$. For $\hat{F_i}({\bf \Theta},D) <   
F_{i}$,
the likelihood is constant, so that all models which predict fluxes less than the
measured upper limit are equally likely.

\subsection{Diffuse Flux Measurements}

To determine the likelihood for model ${\bf \Theta}$ given the measured diffuse
flux in the $i$th beam, we calculate the population-averaged pulsar luminosity for
model parameters $\alpha$, $\beta$, $\gamma$, $n$, $P_0$, and $\b12$
from Eq.~\ref{eq:avglum}, and then use Eq.~\ref{eq:beamflux} to
calculate $\hat{F_i}({\bf \Theta})$, the total predicted flux
from all pulsars in the
beam. We compare this to the measured diffuse flux in the beam
$F_i$, by allowing
the pulsars to comprise
a maximum fraction $\epsilon_{d}$ of the total measured diffuse flux.
Choosing a flat function with a
one-sided Gaussian roll-off with width determined by ${\sigma_{F}}_{i}$, the error on the
measured diffuse flux in beam $i$, as our likelihood function, the PDF 
for model ${\bf \Theta}$ given diffuse measurement $i$ is
\be
{\cal L}_{dif,i}({\bf \Theta}) = \frac{\int_{\hat{F_i}({\bf \Theta})}^{\infty} dF~g(F - \epsilon_{d} F_i);{\sigma_{F}}_{i})}{\int_{\epsilon_{d} F_i}^{\infty} dF~g(F - \epsilon_{d} F_i);{\sigma_{F}}_{i})}
\label{eq:diflike}
\ee
for $\hat{F_i}({\bf \Theta}) \ge \epsilon_{d} F_{i}$. For $\hat{F_i}({\bf \Theta}) < \epsilon_{d} F_{i}$, the likelihood is constant.

The forms of the likelihood functions for detections, upper limits, and diffuse background
measurements
are illustrated in Figure 1.

\subsection{Grid Search and Marginalization}

To find the parameters which best describe the GRP population,
we calculate the total likelihood (Eq.~\ref{eq:totlike}) across a wide range of
values for the parameters $\alpha$, $\beta$, $\gamma$, $n$, $P_0$, and $\b12$
and find that combination
of parameters which maximizes the total likelihood. For each
parameter, we calculate marginalized PDFs as
\be
f_{\Theta_j}(\Theta_j) = \frac{\int_{{\rm exc}~\Theta_j} d{\bf \Theta}~{\cal L}({\bf \Theta})}{\int d{\bf \Theta}~{\cal L}({\bf \Theta})}, 
\label{eq:margpdf}
\ee
so that, for a given parameter $\Theta_j$, the marginalized PDF
is the normalized integral
over all other parameters. A confidence interval $C$ on parameter $\Theta_j$
can be 
calculated from Eq.~\ref{eq:margpdf}
by finding the range of $\Theta_j$ in which a percentage $C$ of the
total probability is contained.

\section{Analysis and Results}

We make several assumptions about the GRP population at the start of the analysis.
For both the OSSE and the EGRET pulsars, we
adopt a spatial distribution (as defined by Eq.~\ref{eq:spatial}) parametrized by
$r_g = 6$ kpc, 
$r_r = 4$ kpc, $w_r = 1.5$ kpc, and $\eta = 0.25$ (n.b. $\eta = 0.25$ corresponds to a 40\% increase
of pulsar number density at the radius of the molecular ring from its value for a Gaussian disk alone). 
For the OSSE pulsars, all of
which have ages less than $10^4$ years, we adopt a scale height of $h = 0.1$ kpc. For the
EGRET pulsars, the oldest of which is $10^{6}$ years old,  we adopt a scale height of
$h = 0.5$ kpc, as these older pulsars will have moved further away from their birth sites.
We take $R_{\odot}$ to be 8.5 kpc throughout.
The corresponding PDF for distance $D$
from the Sun, calculated from Eq.~\ref{eq:distancepdf}, is illustrated in Figure 2. 
We take the age of the Galaxy to be $10^{10}$ years
and take the pulsar birthrate to be 1 per 100 years. We set $\epsgam$, the maximum fraction of spin-down luminosity
a pulsar may convert into gamma-ray luminosity, to 0.5, consistent with the highest $\epsgam$ measured
for the known pulsars ($\epsgam \sim 0.3$ for PSR B1055-52).
Finally, we set $\epsilon_{d}$,
the maximum fraction of the diffuse flux attributable to
pulsars, equal to 0.5. In Sections 5.1 and 5.2, we discuss the effects that changing these
assumptions have on the analysis.

\subsection{OSSE Results}

Using the 3 OSSE detections,
26 upper limits, and 3 diffuse measurements described
in Section 2.1, we
calculate the total likelihood across a wide range of values
for $\alpha$, $\beta$, $\gamma$, $n$, $P_0$, and $\b12$.
The likelihood maximizes for one combination of parameters $\alpha$, $\beta$, and $\gamma$,
and is a maximum across a range of values for $n$, $P_0$, and $\b12$.
Figure 3 shows contours of equal log likelihood as a
function of $\alpha$, $\beta$, and log $\gamma$, taken in pairs.
Figure 4 shows the marginalized PDFs for all 6 parameters. Best parameter values
for $\alpha$, $\beta$, and log $\gamma$, with
95\% confidence intervals calculated from the marginalized PDFs, are
$8.3_{-1.2}^{+1.6}$, $7.6_{-1.1}^{+1.9}$, and $19.4_{-3.6}^{+2.0}$, respectively,
leading to a best-fit luminosity law of
\be
L = 10^{19.4} P^{-8.3} \b12^{7.6}~{\rm ergs/s}.
\label{eq:osselaw}
\ee
The marginalized PDFs allow us to place
a 95\% confidence upper limit on $n$ of 4.75, a
95\% confidence lower limit on $P_0$ of 6.5 ms, and a 95\% confidence upper limit on 
$B$ of $10^{14}$ Gauss. Changing $\epsilon_{d}$,  the fraction of the total diffuse flux that is allowed
to be attributable to pulsars, from our assumed value of 0.5 alters these limits slightly.
Allowing pulsars to
comprise up to 100\% of the total diffuse flux results in a lower limit of
4.5 ms on $P_0$, while allowing pulsars to comprise no more than
10\% of the total diffuse flux leads to an upper limit of 3.75 on $n$. 
In Figure 5, we show how the fraction of OSSE diffuse flux
due to pulsars varies with $n$, $P_0$, and $\b12$. For best-fit values of $\alpha$, $\beta$,
and $\gamma$, the population-averaged pulsar luminosity
(Eq.~\ref{eq:avglum}), and hence the diffuse flux from all pulsars,
 increases with increasing $n$, decreasing $P_0$, and increasing $\b12$. 
Figure 5 shows that unidentified
pulsars may be quite important in contributing to the OSSE diffuse background
and should be considered in future models of diffuse Galactic emission.

Given our best-fit luminosity law and assuming values for $n$, $P_0$, and $\b12$
of 2.5, 10 ms, and $2.5 \times 10^{12}$ Gauss, respectively, we use Eq.~\ref{eq:hist} to
calculate the distribution
of predicted OSSE flux for our model Galactic GRP population.
According to this histogram,
shown in Figure 6, and given OSSE's point source sensitivity of $6.3 \times 10^{31}$ ergs/s 
(in the 50$-$200 keV energy range, and for an integration time of $5 \times 10^{5}$ seconds
(\cite{geh93})),
OSSE could have detected 20
GRPs as steady sources in a full-sky survey.
In Figure 6, we also plot the model-predicted OSSE
fluxes 
for all known spin-driven pulsars with measured period and period derivative
(includes 546 pulsars (\cite{taylor932})).
The pulsar with the fourth highest
predicted flux in the OSSE energy range is PSR B0540$-$69,
the Crab-like pulsar in the
Large Magellanic Cloud at a distance of 50 kpc.
This illustrates the enormous capability a more
sensitive low-energy gamma-ray instrument would have
for detecting young, distant pulsars that are unlikely to be discovered through radio
searches\footnote{We note that 10 hours of integration are necessary to detect B0540$-$69
in the radio (\cite{man93}).}.
 
In
Figure 7, we list the 50 spin-driven pulsars with highest predicted OSSE fluxes in
order of decreasing predicted flux,
along with measured fluxes
or upper limits. Error bars are derived from the uncertainties on both the luminosity law parameters $\alpha$, $\beta$, and $\gamma$
and on the pulsar's
distance. This plot allows us to predict which pulsars are good candidates
to search for gamma-ray emission with more sensitive low-energy gamma-ray
instruments. Because our analysis assumes that pulsars' beams point
directly at us,
a pulsar's actual detectable gamma-ray flux may be lower
than that listed in Figure 7, depending on the orientation of the beam and our line
of sight. However, if the successfully detected pulsars have beams that are somewhat skewed from our line
of 
sight, then the actual flux of other pulsars may be larger
than the prediction. 
  
\subsection{EGRET Results}

Using the 7 EGRET pulsar detections, 353 upper limits, and 3 diffuse background measurements detailed in Section 2.2,
we calculate the likelihood for a wide range of parameter values and again find a well-defined
likelihood maximum in $\alpha$, $\beta$, and $\gamma$.
Figure 8 shows contours of log likelihood vs. pairs of these parameters. The best-fit luminosity law is
\be
L = 10^{32.0} P^{-1.8} \b12^{1.5}~{\rm ergs/s},
\label{eq:egretlaw}
\ee
quite different from the OSSE best-fit law.
Figure 9 shows the marginalized PDFs for all 6 model parameters. Best parameter values for
$\alpha$, $\beta$, and log $\gamma$, with 95\% confidence intervals calculated from
the marginalized PDFs, are
$1.82^{+0.13}_{-0.11}$, $1.54^{+0.13}_{-0.15}$, and $32.04^{+0.07}_{-0.15}$, respectively.
We can place an
upper limit at 95\% confidence on $n$ of 3.6,
a 95\% confidence lower limit
on $P_0$ of 7.1 ms
and a 95\% confidence lower limit on $\b12$ of $1.9\times 10^{10}$ Gauss.

In Figure 10, we show the fraction of EGRET diffuse flux
due to pulsars for our best-fit model
and for various values of $n$, $P_0$, and $\b12$. Because the dependence of the population-averaged luminosity 
(Eq.~\ref{eq:avglum})
on period and magnetic
field is much different than for the OSSE pulsars, the total diffuse EGRET flux varies differently with these
parameters and actually {\it increases} with increasing $P_0$ and {\it decreases} with increasing $\b12$ for $n=2.5$.
Figure 10 show that the 
contribution from unidentified pulsars to the diffuse flux is, at most, a few percent,
and therefore is not important for models of the diffuse background. There are, however, several factors which 
influence this conclusion. For instance, as noted in Section 2.2, if the spectral index
of the diffuse emission is $-1.5$ and not $-2$, integrated diffuse fluxes
would be 3 times greater. In this case, pulsars might comprise as much as 10\% of the diffuse
Galactic emission. Additionally, if the GRP scale height is
lower than the 0.5 kpc that we have assumed,
a greater
fraction of the diffuse flux could be due to pulsars. However, even with a scale height of 0.1 kpc,  
as used for the OSSE population, this fraction is always lower
than 10\% for reasonable values of $n$, $P_0$, and $\b12$. 

In Figure 11, we show the predicted flux distribution for a model population of GRPS given our best-fit luminosity
law (Eq.~\ref{eq:hist}) and for values of $n$, $P_0$,
and $\b12$ of 2.5, 15 ms, and $10^{12}$ Gauss, respectively.
Predicted fluxes for all known radio pulsars with measured $P$ and $\dot{P}$
are also shown.
The point source sensitivity of EGRET ($S_{\rm E} = 1\times10^{-7}$ ph cm$^{-2}$ s$^{-1}$
 = $10^{31.54}$
ergs kpc$^{-2}$ s$^{-1}$) and the expected sensitivity of GLAST
($S_{\rm G} = 4\times10^{-9}$ ph cm$^{-2}$ s$^{-1}$ = $10^{31.54}$
ergs kpc$^{-2}$ s$^{-1}$) are marked by dashed lines. We take $S_{\rm G}$ to be
twice that calculated by Gehrels \& Michelson (1999) for a 2-year all sky survey,
recognizing that because pulsars are primarily a low-latitude population, surveys for them
will suffer from 
the increased diffuse background in the Galactic plane.
We estimate that EGRET should have
detected roughly 20 pulsars. This number
is tantalizing as, of the 73 unidentified EGRET sources within 10 degrees
of the Galactic plane, 17 have non-variable fluxes, as we expect for pulsars (\cite{mcl96}; \cite{mcl992}).
We therefore suggest, as have others, that these 
unidentified EGRET sources are GRPs.
Unfortunately, due to the long integration times needed to detect only a small number of gamma-ray photons from
these sources, searches for pulsars of unknown period in the EGRET data are prohibitive.
Searching for radio pulsar counterparts will help to determine the nature of some
of these sources, but because radio pulsar beams are evidently narrower than gamma-ray
beams, many of these pulsars will be radio-quiet. Therefore, detecting these EGRET
unidentified sources as pulsars will likely have to wait for a more sensitive
high-energy instrument which will allow blind periodicity searches.

According to Figure 11,
GLAST should detect roughly 750 GRPs as steady sources. 
We estimate that 140 of these pulsars (including all unidentified EGRET sources, if they
are indeed GRPs) could be detected through a blind periodicity 
search of GLAST data\footnote{We assume a one month integration, a duty cycle of 1/2, and require
a minimum $N_{s}^{2}/N_{t}$ of
50, where $N_{s}$ is the number of pulsed source counts and $N_t$ is the number of
total counts, for detection, as in Mattox et al. (1996).}.
We note that, using the same detection criterion, 
only 3 EGRET pulsars are detectable through a blind
periodicity search of EGRET data.
Figure 11 also shows that about 120 of the approximately 1000 currently known
radio pulsars should be 
seen by GLAST. With this large sample
of radio and gamma-ray emitting pulsars, which will undoubtedly increase as the Parkes
 survey and other surveys
near completion, one may compare radio and gamma-ray
pulse profiles, 
enabling us to determine where in 
the magnetosphere the gamma-ray emission is coming from and providing
important insights
on the merits of outer gap and polar cap models. For the pulsars which are not 
detectable in the radio and are too faint to perform a blind search with GLAST data,
x-ray observations may yield counterparts. As the angular resolution
of GLAST will
be better than that of EGRET, these searches should
be much more feasible because fewer trial sky positions need to be considered in referencing
photon arrival times to the solar system barycenter.

Our prediction of 750 GLAST-detectable pulsars is dependent upon several assumptions. We
are taking $\epsgam$, the maximum fraction of spin-down luminosity a pulsar may convert to gamma-ray luminosity, to
be 0.5. If the actual value of $\epsgam$ is lower than this, fewer pulsars will be detectable by GLAST. For example, if $\epsgam$ is
actually 0.1, we expect 610 GLAST-detectable pulsars. In
Figure 11, we plot the predicted flux distribution for a population with $\epsgam$ of 0.01. While this value is so low
as to be inconsistent with the measured gamma-ray luminosities of the known EGRET pulsars, the figure illustrates the effect a
lower $\epsgam$ has on the expected flux distribution. We note that, because the scaling of the best-fit OSSE luminosity law with
$P$ and $\b12$ is far steeper than the scaling of $\dot{E}$ with these values, the predicted OSSE flux distribution is 
insensitive to
$\epsgam$.

Our GLAST prediction is also dependent upon 
assuming a birthrate
of 1/(100 yr). It is possible, however, that the actual pulsar birthrate is substantially higher or lower than this value.
Adopting a much higher or lower birthrate while keeping all other parameters fixed would cause the histogram in Figure 11 to
be inconsistent with the EGRET pulsar detections. However, because the flux histogram of Figure
11
depends not
only on the number of pulsars but also on the average luminosity of a pulsar, changing the braking index, magnetic field, 
and initial
spin period can compensate for a reduced birthrate.
For reasonable values of these parameters (i.e. $2 < n < 3$; 1 ms $< P_0 < 30$ ms; $1 < B_{12} < 5$),
birthrates as low as
1/(500 yr) and as high as 1/(25 yr) can be made consistent with the observed GRP flux distribution. 

Our histogram of predicted fluxes also
depends on the spatial distribution model we choose for the pulsar population. Our model does not account for the location of
the spiral arms or for local
density variations, which is evidenced by the discrepancy between our model's predictions and the detections of the bright
Vela, Geminga, and Crab pulsars. Adopting a model with a larger radial scale and/or including spiral arms would cause
the number of GLAST-detectable pulsars and the expected density of pulsars in our region of the Galaxy to increase.

Furthermore, and perhaps most significantly, because GLAST has not yet been built, its 
actual sensitivity is not yet known. Additionally, this sensitivity will be latitude dependent
and will also depend upon models for the diffuse Galactic background. While we defer a more
thorough analysis which includes 
these effects to a later paper, we do find that the number of detectable pulsars varies
with GLAST sensitivity as $S_{\rm G}^{-1}$.
Therefore, if the actual sensitivity is half our 
assumed value, we
expect roughly 1500 GLAST-detectable pulsars. For a sensitivity twice that we have assumed,
we expect 500 GLAST-detectable pulsars.

Figure 12 lists the known pulsars with the top 50 predicted EGRET fluxes,
along with measured fluxes
and upper limits, indicating which pulsars are good candidates for
gamma-ray detection
with future instruments. Error bars are derived from the uncertainties on $\alpha$, $\beta$, $\gamma$, 
and distance. Again, as discussed in Section 5.1, 
a pulsar's actual detectable gamma-ray
flux may be higher or lower than that listed in Figure 12, depending upon the correctness of
our assumptions about beaming. 
We note
that the measured and predicted fluxes for PSR B1055-52 differ significantly.
Some of this may be due to the uncertain alignment between the rotation axis and the line
of sight
to this pulsar (\cite{lyne88}; \cite{rankin93}).
 There is also evidence that this pulsar
is much closer than we have assumed.
While we derive a distance of 1.5 kpc from its
dispersion measure and the Taylor and Cordes (1993) model, 
\"Ogelman and Finley (1993), in their analysis of the ROSAT x-ray data, found
that a distance of 500 pc would produce
a more realistic estimate of the neutron star radius (15 km compared to 30 km for the
1.5 kpc estimate).
Furthermore, Combi, Romero, and Azc\'arate (1997) derive a distance estimate
of 700 pc from a study of the extended nonthermal radio source around the pulsar.
Therefore, in Figure 12, we also mark the predicted flux for PSR B1055-52 for a revised distance of
500 pc. It is interesting that adopting a distance of 500 pc removes the constraint that this pulsar
must be converting 30\% of its
spin-down energy into observable gamma-ray luminosity. With the revised distance estimate, a conversion percentage of only 3\% is 
necessary. Similarly, we may also be overestimating the distance to Vela.
 While we are using the commonly quoted distance of 0.5 kpc for Vela (\cite{taylor932}),
recent spectroscopic evidence suggests that this pulsar may actually be only 0.25 kpc away
(Cha et al. 1999). Adopting this revised distance would alter our best-fit model modestly, but would have a large effect on
the implied efficiency of gamma-ray production for the Vela pulsar, lowering it from 1.5\% to 0.4\%. Incidentally, assuming a 
0.5 kpc distance for B1055-52, the Geminga pulsar is the most efficient converter of spin-down energy to gamma-rays, with an 
implied efficiency of 22\%. The Crab pulsar, with an implied efficiency of 0.01\%, is the least efficient of the EGRET pulsars.

\section{Discussion}

We have used OSSE and EGRET pulsar detections, measured upper limits, and diffuse background measurements
to constrain the
luminosity law and some properties of the gamma-ray pulsar population.
The luminosity laws are quite different at OSSE and EGRET energies, suggesting that different
mechanisms are responsible for
pulsar low- and high-energy gamma-ray emission. The OSSE law
($L \propto P^{-8.3} \b12^{7.6}$) is roughly
consistent with optical pulsar detections (\cite{pacini83}) and is steeper than that
found for ROSAT/ASCA pulsar detections
(\cite{ogel93}; \cite{becker96}; \cite{saito97}).
It is quite different from models
which predict that a pulsar's gamma-ray luminosity should be some fraction of the total spin-down
luminosity (i.e. $L \propto P^{-4} \b12^{2}$). The EGRET law 
($L \propto P^{-1.8} \b12^{1.5}$) is more similar to this 
spin-down
law and even more consistent with a luminosity law
which scales with 
the open-field line voltage drop (i.e. $L \propto P^{-2} \b12$).
It is consistent with the basic polar cap
(\cite{dau96}) and outer gap (\cite{zhang97}) models that have been published.
More gamma-ray pulsar detections and a 
comparison between radio and gamma-ray pulsar beam shapes and properties are 
necessary to discriminate between models and the many versions of them.

We note that a pulsar's age is proportional to $P^{2} \b12^{-2}$, so
that the gamma-ray
efficiency, or the fraction of total spin-down energy converted to 
gamma-ray luminosity, {\it decreases} with increasing age for OSSE pulsars and
{\it increases} with increasing age for EGRET pulsars. Because
both the OSSE and EGRET best-fit luminosity laws differ from the standard $\dot{E}$ model,
they produce a new ranking of pulsars as candidates for gamma-ray detection
and result in new predictions for the pulsar population we expect to see
with future more sensitive gamma-ray instruments.

We find that a more sensitive low-energy gamma-ray
instrument (50$-$200 keV energy range)
would detect young, distant pulsars which are most likely not 
detectable through their radio emission. This would be enormously helpful for
studies of young pulsars, 
pulsar populations, and
supernova physics. A more sensitive high-energy gamma-ray instrument 
(0.1$-$100 GeV energy range)
 such as GLAST would detect
roughly 750 GRPs as steady sources, most of which will have radio beams
misaligned from our line of sight. Of these 750 GRPs, roughly 120 will be known radio pulsars.
About 140 of the 750 GRPs will be detectable as pulsars
through blind periodicity searches, giving us a much more complete picture of the
GRP population. GLAST
will be able to do blind
periodicity searches on all of the unidentified EGRET sources, of
which 20 are probably GRPs. We note that all predictions are dependent upon an assumed $\epsgam$ and
birthrate and will vary for different values of these parameters. While
there is currently a large range of acceptable values for
these parameters, comparing model predictions and actual GLAST detections will enable us to
constrain these parameters much more 
tightly. Our prediction will also obviously depend upon the final sensitivity of GLAST, with the
number of detectable pulsars scaling roughly as $S_{\rm G}^{-1}$.

We estimate the
contribution of pulsars to the measured diffuse background in the OSSE and EGRET
energy ranges, and find that while pulsars may be an important component of
the diffuse
background at OSSE energies, they are most likely not important at EGRET energies.
We are able to use measurements of the diffuse background
to constrain the
braking index, initial spin period, and magnetic field of the pulsar population. 
The OSSE background allows us to place a 95\% confidence
upper limit
of 4.75 on the pulsar braking index, a lower limit of 6.5 ms on the pulsar
 initial
spin period, and an upper limit of $10^{14}$ Gauss on the pulsar
magnetic
field, while the EGRET background implies a 95\% confidence upper limit of 3.6 on $n$, 
a lower limit of 7.1 ms on $P_0$, and a lower limit of $1.9 \times 10^{10}$ Gauss on B.
We note that there are large errors on these parameters, and that the real
distributions of these quantities are likely to be more complicated than we have assumed.
Again, GLAST should allow us to constrain these parameters far more tightly than currently possible.

In order to do a more sophisticated analysis of the GRP population, more GRP detections are
necessary. Large strides in our understanding of GRPs will likely
have to wait until the launch of GLAST. In the meantime, however, there are useful
things that can be done to further our understanding. While searches for
periodicities in EGRET data are very difficult, they may not be impossible if one
takes an iterative approach (i.e. searching small, sequential intervals
of data sets where fewer
trial periods and positions are necessary) 
 and uses Bayesian methods to detect pulses of unknown shape
 (e.g. \cite{mcl99}).
 Searches for radio counterparts of the
unidentified EGRET sources likely to be GRPs are also important, as some
GRPs will emit in the radio and even non-detections provide
useful information. Furthermore, whenever new x-ray or radio pulsars are detected, especially
young or nearby pulsars, one should return to the OSSE and EGRET data and fold them.
An accurate
ephemeris may not be available but a known period allows a smaller range of parameters to
search and a reduction of the search threshold.
Some of the Parkes
multibeam pulsars (\cite{camilo99}), for instance,
are excellent gamma-ray candidates.

\acknowledgments

We thank M. Ulmer for providing OSSE upper limits and for useful discussions. We
thank Z. Arzoumanian for providing updated EGRET upper limits, for useful discussions,
and for a careful reading of the manuscript. We also thank D. Helfand and D. Thompson for 
thought provoking discussions, and the anonymous referee for useful comments.
M. A. McLaughlin acknowledges support from the
NASA/NY Space Grant Program. The research was supported by NSF Grants AST-9528394 and
AST-9819931
and NASA Grant NAG5-3515 to Cornell University.

{}

\newpage

\begin{figure}[t]
\plotfiddle{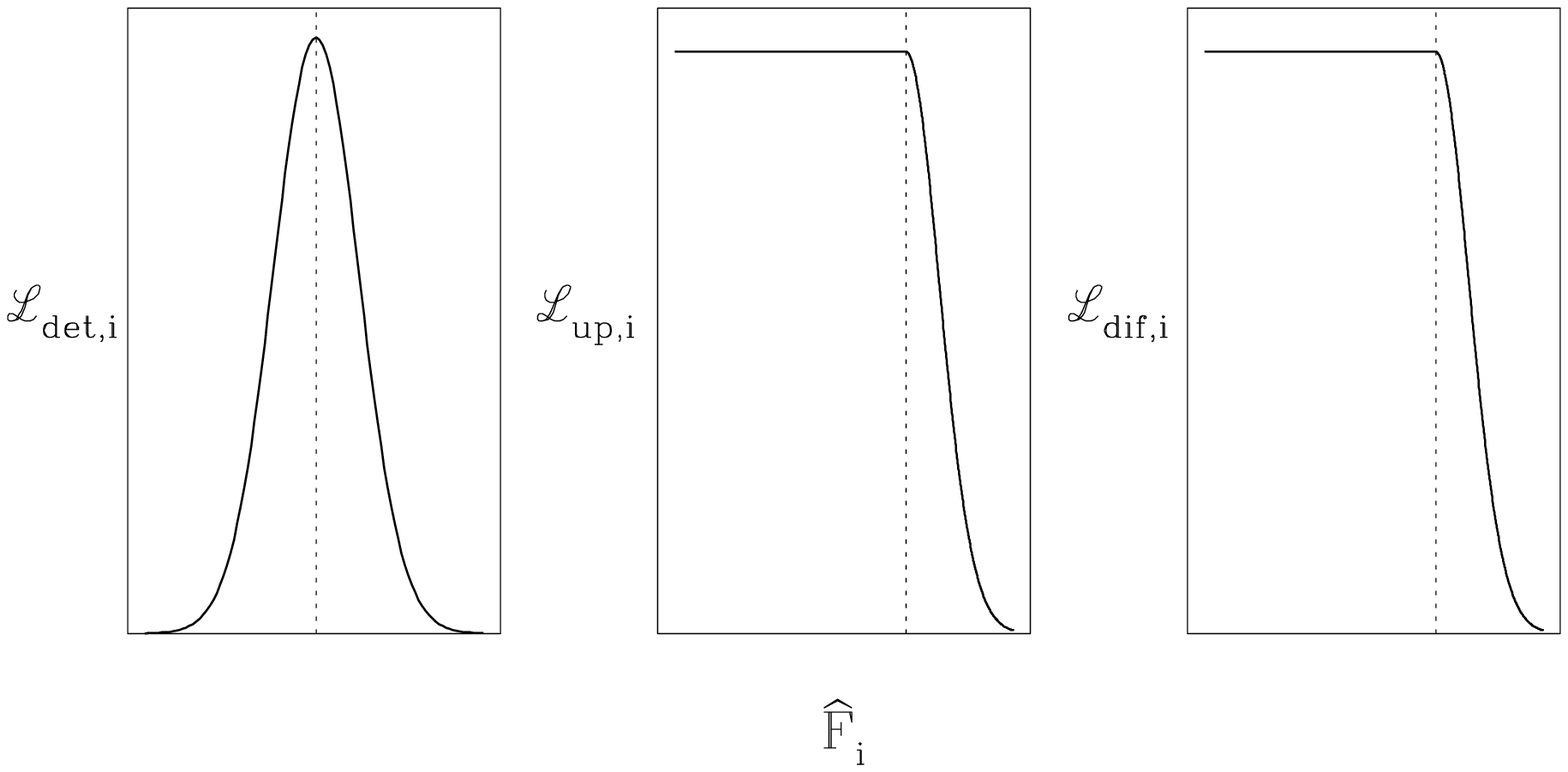}{2.0truein}{0}{60}{60}{-200}{-100}
\caption{Likelihood functions for detections, upper limits, and diffuse
measurements are plotted against predicted model flux, $\hat{F_i}$. Dotted
lines mark where predicted flux $\hat{F_i}$ equals the measured detection,
upper limit, or diffuse measurement.}
\end{figure}
\begin{figure}[b]
\plotfiddle{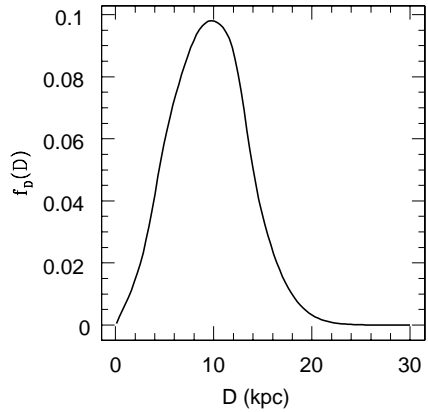}{2.0truein}{0}{60}{60}{-100}{-70}
\caption{The PDF for distance D from the Sun for a model population of GRPs, calculated for
a model parametrized by $r_g = 6$ kpc, $h = 0.1$ kpc,
$r_r = 4$ kpc, $w_r = 1.5$ kpc, and $\eta = 0.25$, and for $R_{\odot}$ = 8.5 kpc.}
\end{figure}
\begin{figure}[h]
\centerline{\psfig{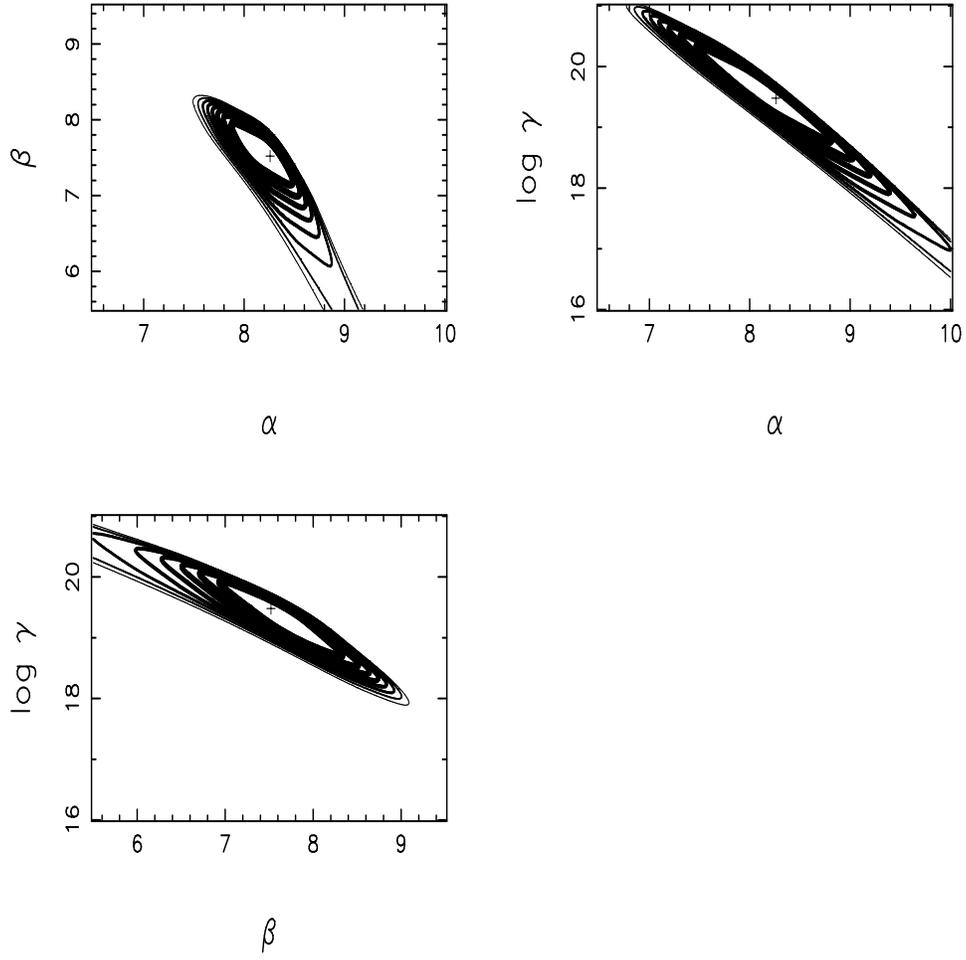}}
\caption{Contours of equal log likelihood vs. pairs of parameters for the OSSE data.
Crosses mark the maximum of the log likelihood.}
\end{figure}
\begin{figure}[h]
\centerline{\psfig{figure=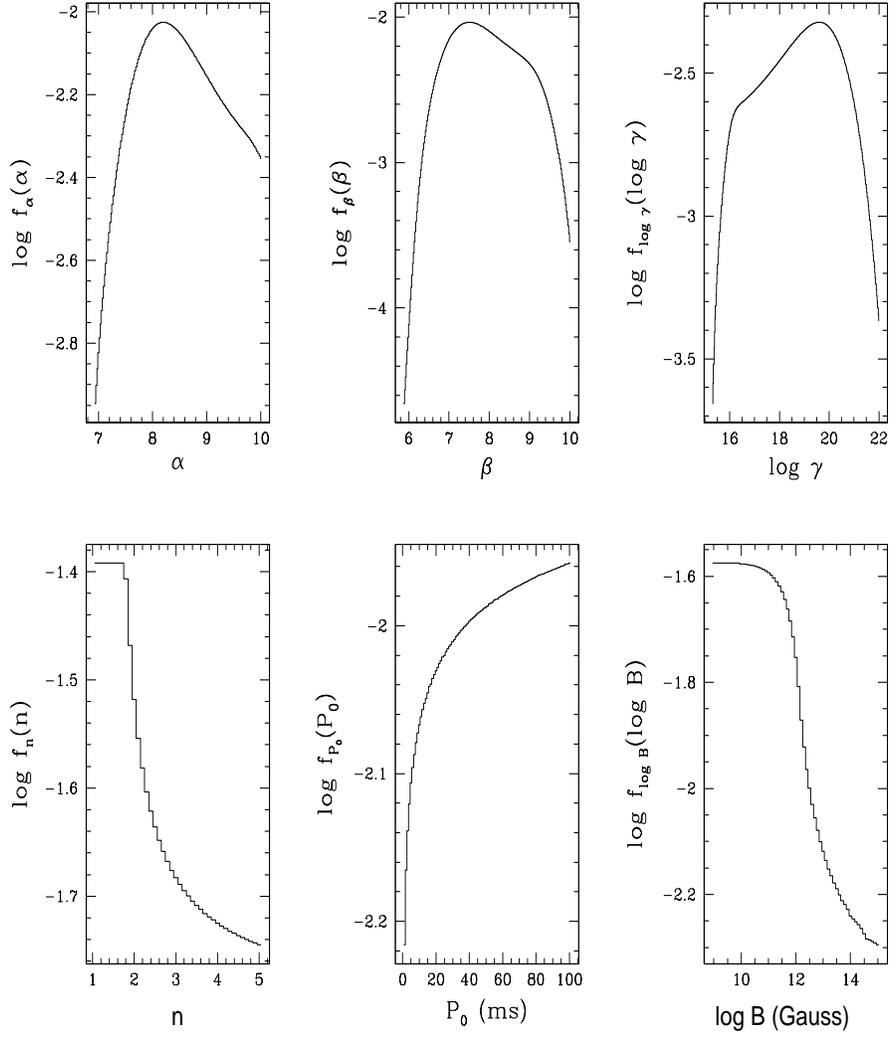,height=6.0in,width=5.0in}}
\caption{Marginalized PDFs for all parameters for the OSSE
data.}
\end{figure}
\begin{figure}[h]
\plotfiddle{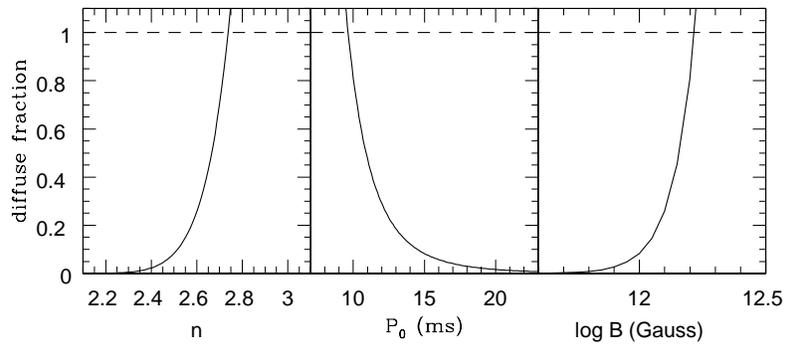}{2.0truein}{0}{60}{60}{-200}{-100}
\caption{The fraction of OSSE diffuse flux attributable to unidentified pulsars
as a function of braking index $n$, initial spin period $P_0$, and surface
magnetic field strength $\b12$.
One parameter is varied and other parameters are kept constant at
nominal values of
$n = 2.5$, $P_0 = 15$ ms, and $\b12 = 1.0$. The dashed line shows where the fraction of diffuse flux
attributable to pulsars reaches unity and defines a range of parameter values which are disallowed.}
\end{figure}\begin{figure}[h]
\centerline{\psfig{figure=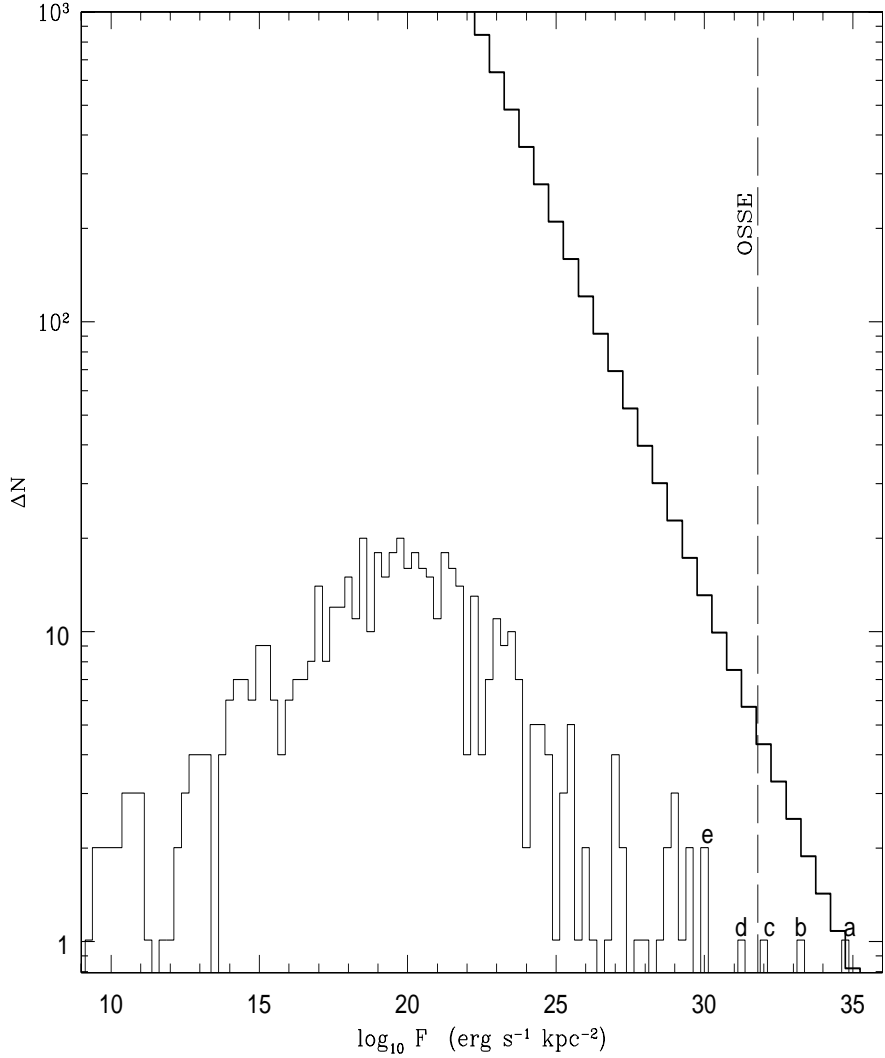,height=6.0in,width=5.0in}}
\caption{Histogram of predicted OSSE pulsar flux for $\alpha$ = 8.3, $\beta$ = 7.6, log
$\gamma$ =
19.4, $n$ = 2.5, $P_0$ = 10 ms, and $\b12$ = 2.5. The thick solid line shows the predicted
 flux distribution for a model population of GRPs. The dashed line denotes the point source sensitivity of OSSE.
The thin solid line shows
the predicted fluxes for known pulsars with measured $P$ and $\dot{P}$. 
Bins with highest predicted fluxes are labeled and correspond to the following pulsars: a) Crab; b) B1509-59; c) Vela;
d) B0540-69; e) B1610-50, B1706-44.}
\end{figure}
\begin{figure}[h]
\centerline{\psfig{figure=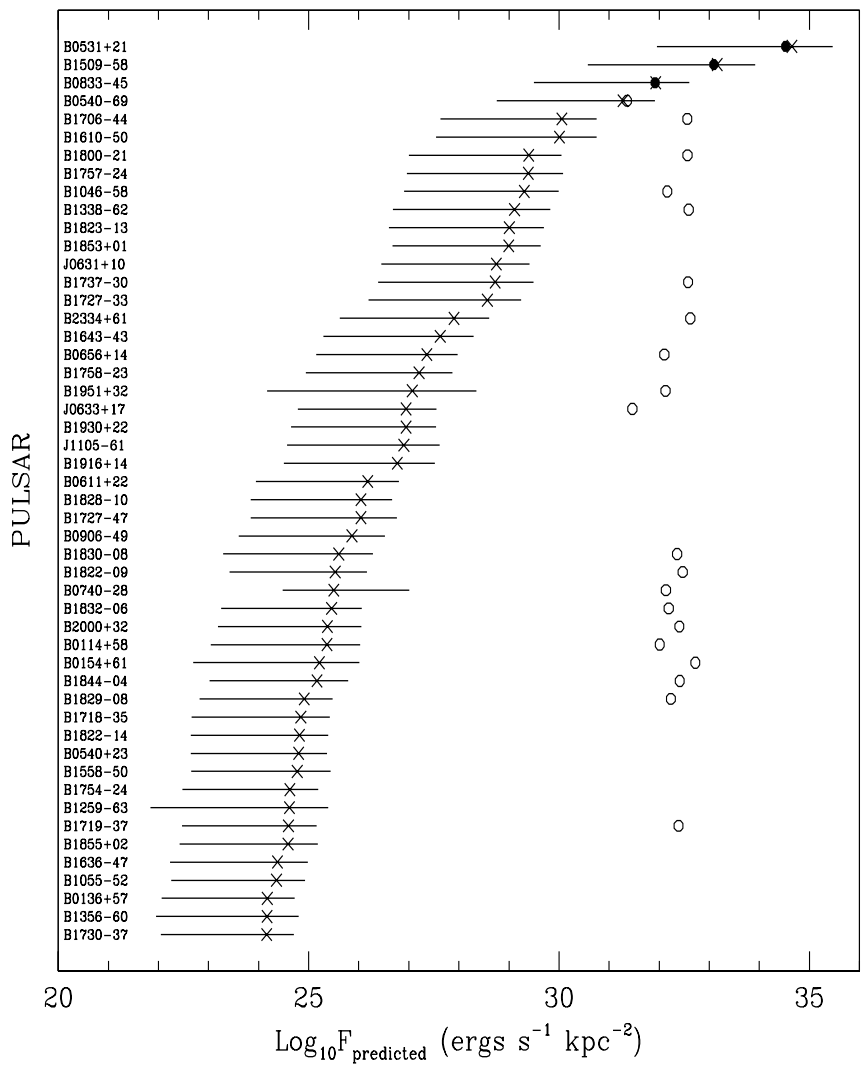,height=6.0in,width=5.0in}}
\caption{Known radio pulsars listed in order of decreasing predicted OSSE flux.
Crosses denote predicted gamma-ray fluxes, with error bars.
Solid dots mark OSSE-detected fluxes, while open circles indicate upper limits.
If the distance to B1055$-$52 is indeed 500 pc, instead of the assumed 1.5 kpc,
this pulsar's predicted flux would be 9 times higher, making it the 35th (as opposed to
47th) entry in this table.}
\end{figure}
\begin{figure}[h]
\centerline{\psfig{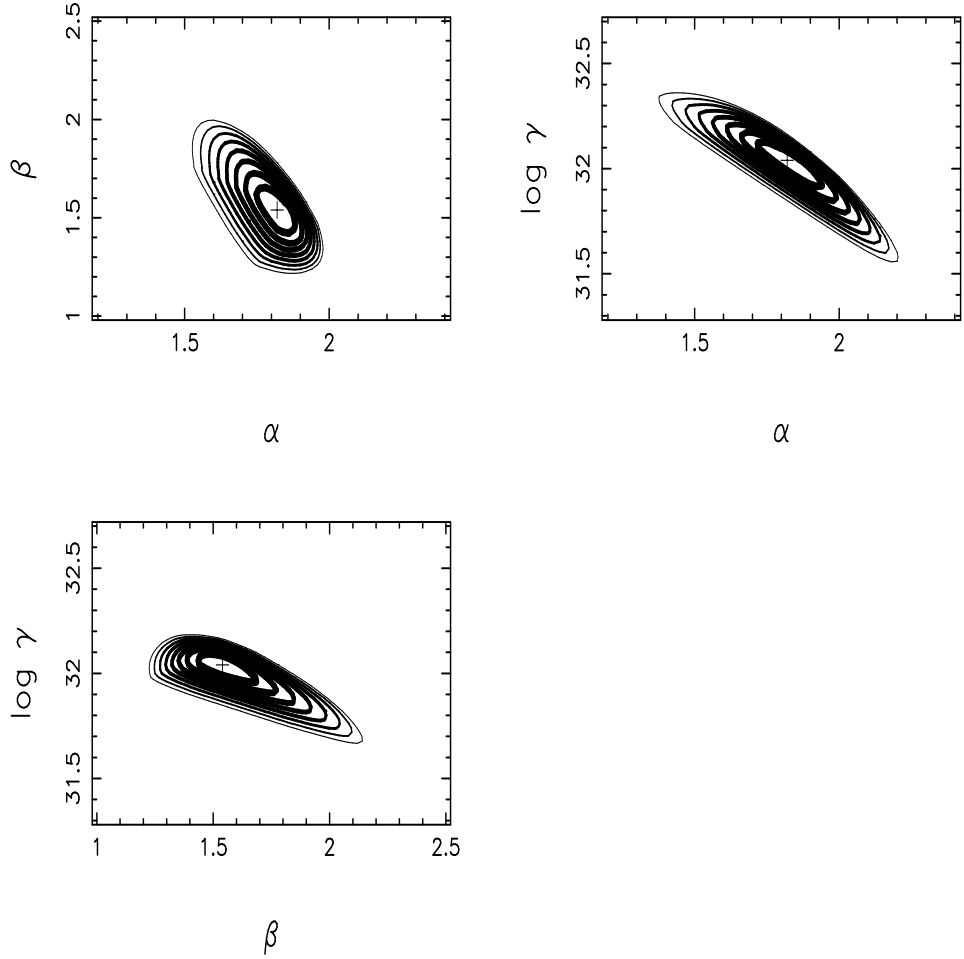}}
\caption{Contours of equal log likelihood vs. pairs of parameters for EGRET data. Crosses mark the
maximum of the log likelihood.}
\end{figure}
\begin{figure}[h]
\centerline{\psfig{figure=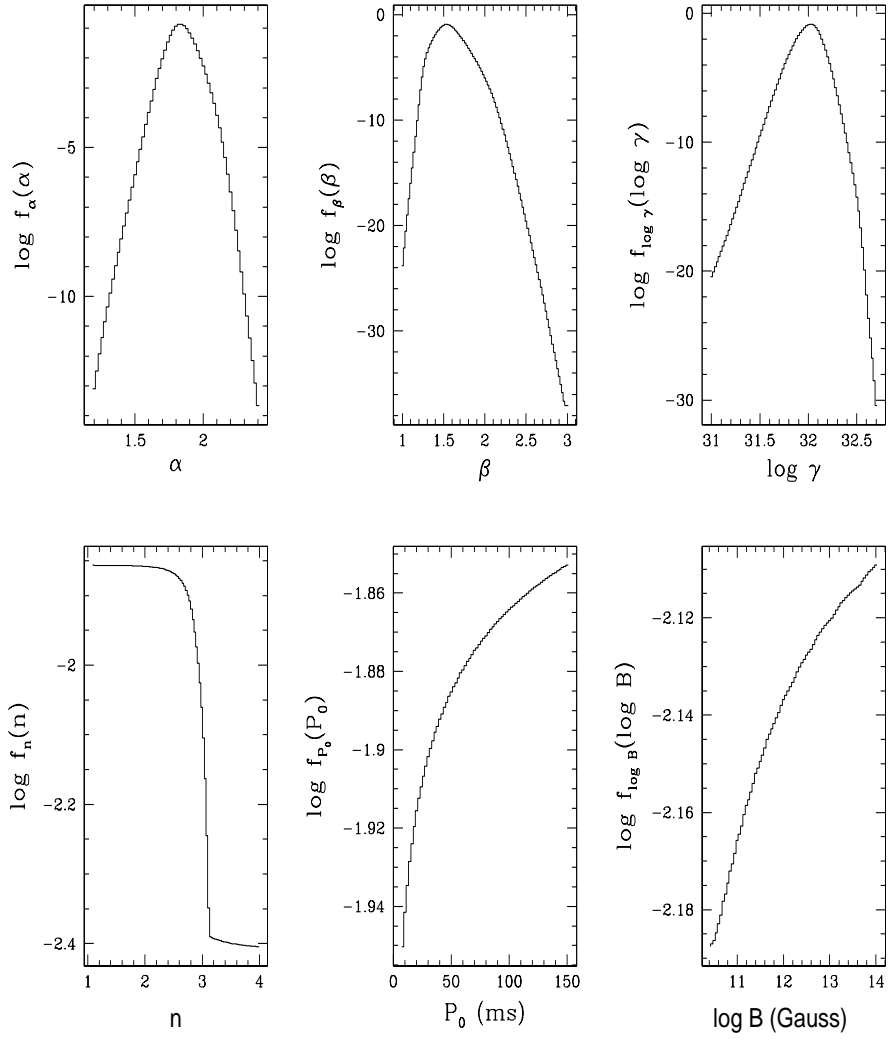,height=6.0in,width=5.0in}}
\caption{Marginalized probability density functions for all parameters for EGRET
data.}
\end{figure}
\begin{figure}[h]
\plotfiddle{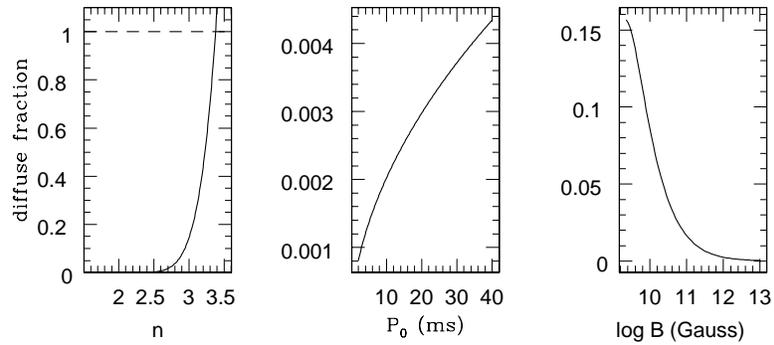}{2.0truein}{0}{60}{60}{-200}{-100}
\caption{The fraction of EGRET diffuse flux attributable to unidentified pulsars
as a function of braking index $n$, initial spin period $P_0$, and surface
magnetic field $\b12$.
As one parameter is varied, other parameters are kept constant at values of
$n = 2.5$, $P_0 = 15$ ms, and $\b12 = 1.0$. The dashed line shows where the fraction of diffuse flux
attributable to pulsars reaches unity and defines a range of disallowed values for $n$.}
\end{figure}
\begin{figure}[h]
\centerline{\psfig{figure=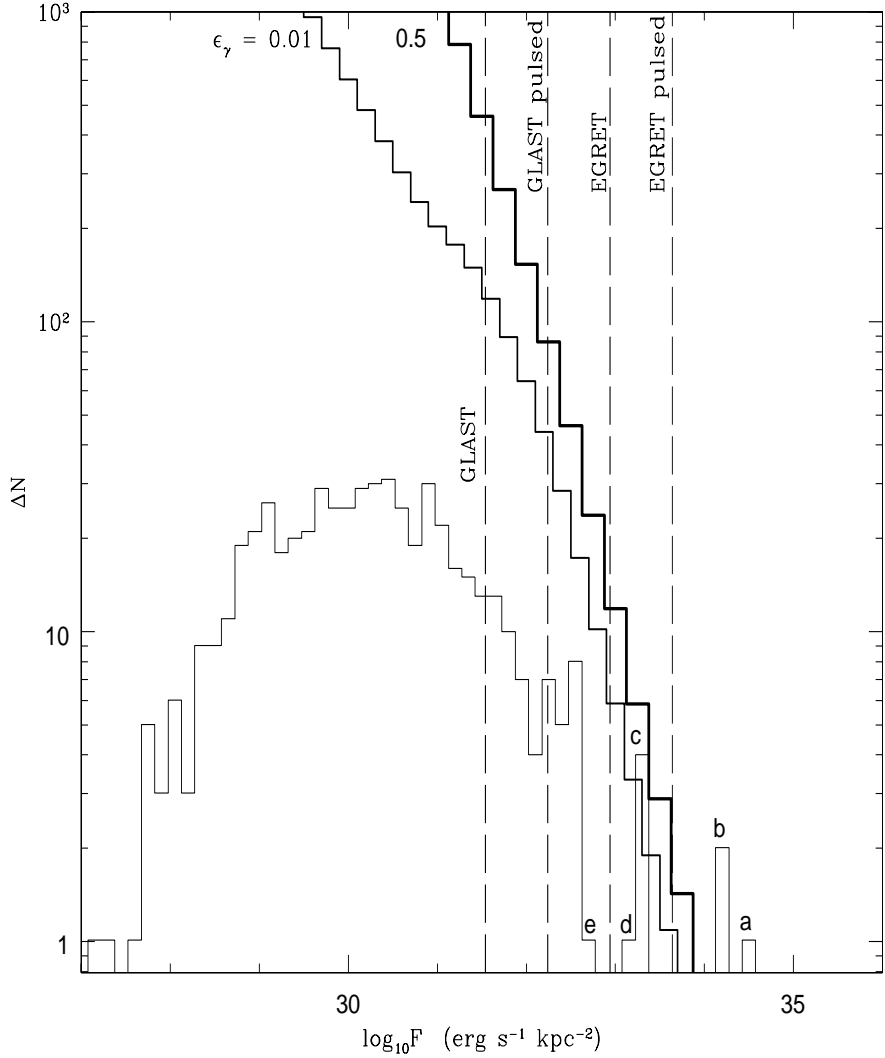,height=6.0in,width=5.0in}}
\caption{Histogram of predicted EGRET pulsar flux for $\alpha$ = 1.8, $\beta$ = 1.5,
log $\gamma$ = 32.0, $n$ = 2.5, $P_0$ = 15 ms, and $\b12$ = 1.0.
The thickest solid line shows the predicted
pulsar flux distribution of a model population of GRPs, assuming $\epsgam$ = 0.5. The thinner line illustrates the
predicted distribution for $\epsgam$ = 0.01. Dashed lines show the sensitivities of
 EGRET and GLAST
to steady sources and for a blind periodicity search. The thin solid line shows
the predicted fluxes for known pulsars.
Bins with highest predicted fluxes are labeled and correspond to the following pulsars:
a) Vela; b) Geminga, Crab; c) J0631+105, B0656+14, B1509-58, B1706-44; d) B0950+08; e) B1046-58.}
\end{figure}
\begin{figure}[h]
\centerline{\psfig{figure=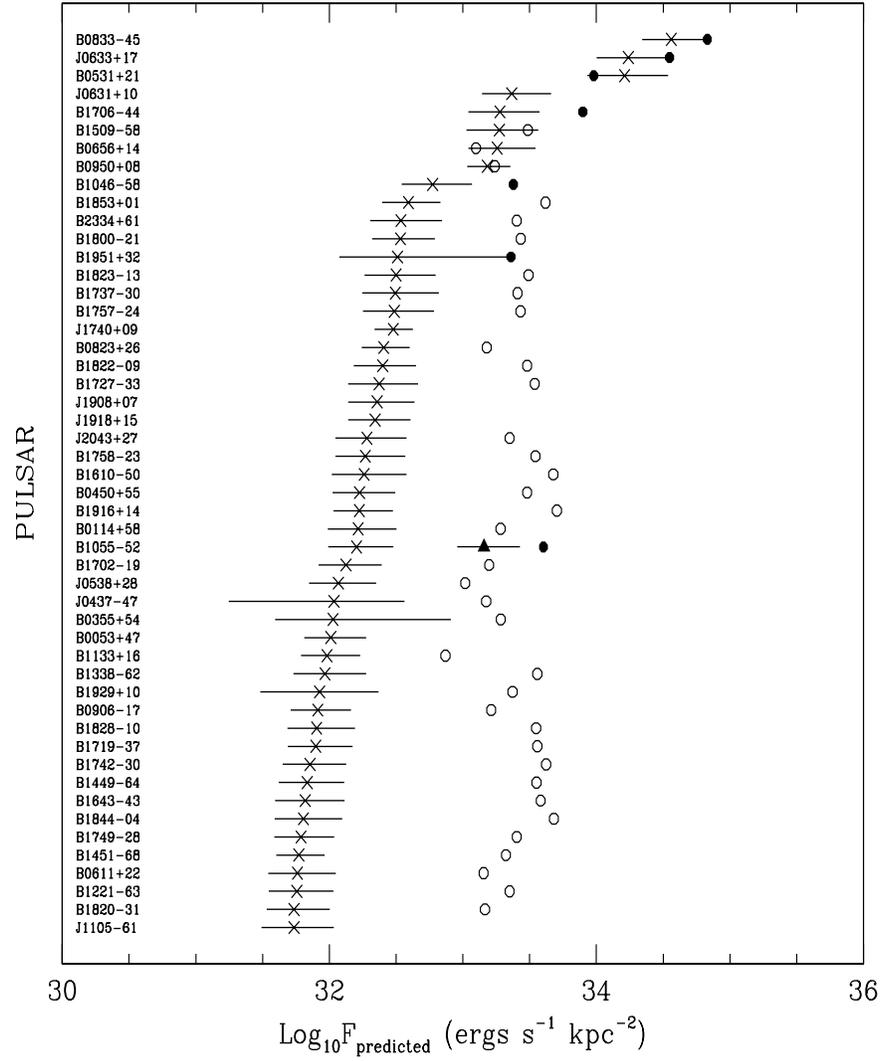,height=6.0in,width=5.0in}}
\caption{Known radio pulsars listed in order of decreasing predicted EGRET flux. 
Crosses denote predicted gamma-ray fluxes, with error bars.
Solid dots mark detected fluxes, while open circles indicate upper limits. The triangle
marks the predicted flux for PSR B1055-52 for a distance of 500 pc.}
\end{figure}


\newpage
\begin{thebibliography}{}
\bibitem[Arzoumanian 1997]{arz97} Arzoumanian, Z., 1997, personal communication
\bibitem[Becker \& Trumper 1996]{becker96} Becker, W. \& Trumper, J., 1996, A\&AS, 120, 69
\bibitem[Camilo et al. 1999]{camilo99} Camilo, F. et al. 1999, ``The Parkes Multibeam Pulsar Survey", IAU Coll. 177, PASP
\bibitem[Cha et al. 1999]{cha99} Cha, A. N., Sembach, K. R., \& Danks, A. C., ApJ, 515, L25-L28
\bibitem[Combi, Romero, \& Azc\'arate 1997]{combi97} Combi, J. A., Romero, G. E., \& Azc\'arate, I. N., 1997, ApSS, 250, 1
\bibitem[Daugherty \& Harding 1996]{dau96} Daugherty, J. K., \& Harding, A. K., 1996, ApJ, 458, 278
\bibitem[Fierro et al. 1995]{fierro95} Fierro, J. M., et al., 1995, ApJ, 447, 807
\bibitem[Gehrels, Chipman, \& Kniffen 1993]{geh93} Gehrels, N., Chipman, E., \& Kniffen, D. A., 1993, ``The Compton Observatory in Perspective", AIP Conference Proceedings 280, New York, NY 
\bibitem[Gehrels \& Michelson 1999]{gehrels99} Gehrels, N., \& Michelson, P., 1999, APh, 11, 277
\bibitem[Gouiffes et al. 1992]{gouiffes92} Gouiffes, C., Finley, J., \& \"Ogelman, H., 1992, ApJ, 394, 581
\bibitem[Groth 1975]{groth75} Groth, E.J., 1975, ApJS, 29, 431
\bibitem[Hunter et al. 1997]{hunter97} Hunter, S. D., et al., 1997, ApJ, 481, 205
\bibitem[Kaspi et al. 1994]{kaspi94} Kaspi, V. M., et al., 1994, ApJ, 422, L83
\bibitem[Kaspi et al. 1999]{kaspi99} Kaspi, V. M., et al. 1999, submitted
\bibitem[Kinzer et al. 1999]{kinzer99} Kinzer, R. L., Purcell, W. R., \& Kurfess, J. D., 1999, ApJ, 515, 215
\bibitem[Kuiper et al. 1999]{kuiper99} Kuiper L., Hermsen W., Verbunt F., et al., 1999, Proc. 3rd INTEGRAL Workshop: ``The Extreme Universe", Taormina, Italy
\bibitem[Lyne et al. 1993]{lyne93} Lyne, A. G., Pritchard, R. S., \& Graham-Smith, F., 1993, MNRAS, 265, 1003
\bibitem[Lyne et al. 1996]{lyne96} Lyne, A. G., Pritchard, R. S., Graham-Smith, F., \& Camilo, F., 1996, {\it Nature}, 381, 497
\bibitem[Lyne \& Manchester 1988]{lyne88} Lyne, A. G., \& Manchester, R. N., 1988, MNRAS, 234, 477
\bibitem[Malofeev \& Malov 1997]{mal97} Malofeev, V. M. \& Malov, O. I., 1997, \it Nature \rm, 389, 697
\bibitem[Manchester et al. 1985]{man85} Manchester, R. N., Newton, L. M., \& Durdin, J. M., 1985, \it Nature\rm, 313, 374
\bibitem[Manchester \& Peterson 1989]{man89} Manchester, R. N. \& Peterson, B. A., 1989, ApJ, 342, 23
\bibitem[Manchester et al. 1993]{man93} Manchester, R. N., et al., 1993, ApJ, 403, L29
\bibitem[Mattox et al. 1996]{mattox96} Mattox, J. R., et al. 1996, A\&AS, 120, 95
\bibitem[Matz et al. 1994]{matz94} Matz, S. M., et al. 1994, ApJ, 434, 288
\bibitem[McLaughlin et al. 1996]{mcl96} McLaughlin, M. A., Mattox, J. R., Cordes, J. M., \& Thompson, D. J., 1996, ApJ, 473, 763
\bibitem[McLaughlin \& Cordes 1999]{mcl992} McLaughlin, M. A., \& Cordes, J. R., 1999, in preparation
\bibitem[McLaughlin \& Cordes 1999]{mcl99} McLaughin, M. A. Cordes, J. M., Hankins, T. H., \& Moffett, D.A., 1999, ApJ, 512, 929 
\bibitem[Nel et al. 1996]{nel96} Nel, H. I., et al., 1996, ApJ, 465, 898
\bibitem[Nolan et al. 1996]{nolan96} Nolan, P. L., et al., 1996, A\&AS, 120, 61
\bibitem[\"Ogelman et al. 1993]{ogel93} \"Ogelman, H. B., et al., 1993, AdSpR, 131, 351
\bibitem[\"Ogelman \& Finley 1993]{ogel93.2} \"Ogelman, H. B. \& Finley, J. P., 1993, ApJ, 413, L31
\bibitem[Pacini \& Salvati 1983]{pacini83} Pacini, F., \& Salvati, M., 1983, ApJ, 274, 369
\bibitem[Rankin 1993]{rankin93} Rankin, J. M., 1993, ApJS, 85, 145
\bibitem[Saito 1997]{saito97} Saito, Y., 1997, PhD. Thesis
\bibitem[Schroeder et al. 1995]{sch95} Schroeder, P. C., et al., 1995, ApJ, 450, 784
\bibitem[Skibo et al. 1997]{skibo97} Skibo, J. G., et a., 1997, ApJ, 483, L95
\bibitem[Strickman et al. 1996]{strickman96} Strickman, M. S., 1996, ApJ, 460, 735
\bibitem[Taylor \& Cordes 1993]{taylor93} Taylor, J. H., \& Cordes, J. M., 1993, ApJ, 411, 674
\bibitem[Taylor, Manchester, \& Lyne 1993]{taylor932} Taylor, J. H., Manchester, R. N., \& Lyne, A. G., 1993, ApJS, 88, 529
\bibitem[Hartman et al. 1999]{hartman99} Hartman, R. C., et al., 1999, ApJS, 123, 79
\bibitem[Ulmer et al. 1991]{ulmer91} Ulmer, M. P., Purcell W. R., Wheaton, W. A., \& Mahoney, W. A., 1991, \apj, 369, 485
\bibitem[Ulmer et al. 1994]{ulmer94} Ulmer, M. P., et al. 1994, ApJ, 432, 228 
\bibitem[Ulmer et al. 1999]{ulmer99} Ulmer, M. P., et al., 1999, in preparation
\bibitem[Wilson et al. 1993]{wilson93} Wilson, R. B., et al. 1993, in Isolated Pulsars, ed. K. van Riper, R. Epstein, \& C. Ho (Cambridge: Cambridge University Press), 257 
\bibitem[Zhang \& Cheng 1997]{zhang97} Zhang, L. \& Cheng, K. S., 1997, ApJ, 487, 370
\end{thebibliography}
\end{document}